\documentclass[12pt]{article}
\usepackage{graphicx}

\newsavebox{\sboxpubnumber}
\newsavebox{\sboxpubdate}

\newcommand{\Title}[1]{\begin{center} {\Large #1 } \end{center}}
\newcommand{\Author}[1]{\begin{center}{ \sc #1} \end{center}}
\newcommand{\Address}[1]{\begin{center}{ \it #1} \end{center}}

\newenvironment{Abstract}{\begin{quotation}  }{\end{quotation}}
\newenvironment{Presented}{\begin{quotation} \begin{center}
             PRESENTED AT\end{center}\bigskip
      \begin{center}\begin{large}}{\end{large}\end{center}
      \end{quotation}}

\begin{document}

%%%%%%%%%%%%%%%%%%%%%%%%%%%%%%%%%%%%%%%%%%%%%%%%%%%%%%%%%%%%%%%%%%%%%%%%
%%
%% START EDITING HERE!
%%
%%%%%%%%%%%%%%%%%%%%%%%%%%%%%%%%%%%%%%%%%%%%%%%%%%%%%%%%%%%%%%%%%%%%%%%%
\begin{titlepage}
%\pubdate{\today}                    %fill in the date
%\pubnumber{XXX-XXXXX \\ YYY-YYYYYY} %preprint number(s)

\vfill
\Title{A measure of gravitational entropy and structure formation}
\vfill
\Author{Manfred P. Leubner}
\Address{Institute for Theoretical Physics, University of Innsbruck\\
         A-6020 Innsbruck, Austria}
\vfill
%\andauth
%\vfill
%\Author{Your Coauthors Name}
%\Address{Department, Institute \\
%         Postal address}
\vfill
\begin{Abstract}
Increasing inhomogeneity due to gravitational clumping reflects increasing
gravitational entropy in a time evolving universe. Starting from an ensemble
of uniformly distributed particles it is demonstrated that gravitational
clustering is subject to a specific quantization rule for the amount of
increase of gravitational entropy during the formation of inhomogeneities.
The gain of gravitational entropy at each higher order merging process
within the system is shown to result as a natural consequence from an
extremal condition involved. The resulting discrete spectrum of nested,
bound structures of specific mass and radius, ranging from the particle
physics scale to galaxies and super clusters, provides a unified view of
fundamental inhomogeneity scales in the universe from gravitational entropy
considerations. Consequently, also the gravitational arrow of time points in
the direction of stepwise increasing entropy or inhomogeneity, respectively.
\end{Abstract}
\vfill
\begin{Presented}
    COSMO-01 \\
    Rovaniemi, Finland, \\
    August 29 -- September 4, 2001
\end{Presented}
\vfill
\end{titlepage}
\def\thefootnote{\fnsymbol{footnote}}
\setcounter{footnote}{0}

%%%%%%%%%%%%%%%%%%%%%%%%%%%%%%%%%%%%%%%%%%%%%%%%%%%%%%%%%%%%%%%%%%%%%%%%
% The document starts here
%%%%%%%%%%%%%%%%%%%%%%%%%%%%%%%%%%%%%%%%%%%%%%%%%%%%%%%%%%%%%%%%%%%%%%%%

\section{Introduction}

The question why we observe in the universe discrete structure scales as
elementary particles, stellar systems, globular clusters or galaxies,
but nothing between, was originally addressed by Chandrasekar \cite%
{Chandrasekar37}. Today we argue that a suitable concept for the
gravitational entropy of a discrete matter distribution is required to
provide an understanding of the formation of structure scales in an
expanding universe \cite{Ellis01}. In contrast to thermodynamic systems
driven to a uniform distribution, the components of gravitating systems tend
to clump, thus implying a gravitational arrow of time, which points in the
direction of growing inhomogeneity. The universe acts as a self-organizing
system evolving spontaneously into increasingly complex structures due to
the long range nature of gravitational interaction. Contrary to concepts
of black hole entropy \cite{Bekenstein01}, and references therein,
only a few attempts have been made to quantify the apparently
contradictory behavior of thermodynamic and gravitating systems with regard
to a consistent measure of gravitational entropy. Penrose \cite{Penrose90}
introduced the Weyl curvature hypothesis in view of a measure of local anisotropy
and a phase space approach was suggested where gravitational entropy is
interpretable as lack of knowledge of the field configuration considered
\cite{Rothman97}. Presently, no unique definition of a gravitational entropy
for ordinary bound systems is available.

The building blocks of matter are atoms which themselfes are made up by
nucleons where those in turn are subject to quarks and gluons as
constituents. As pointed out by Bekenstein \cite{Bekenstein01}
there is a variety of states and entropy available and obviously,
the deeper we look into matter the more degrees of freedom are accessible
and the higher is the entropy. Let me generalize this view from levels
far above of condensed matter and start with superclusters made up by
clusters of galaxies. These structures are found to have globular clusters
and stellar systems as substructures and we ask if there is a link between
all structure scales available, a quantized gravitational entropy formalism
resulting in a sequence of fundamental inhomogeneities \cite{Leubner00a}.   

On observational grounds we can argue that any concept evaluating
gravitational entropy from a monotonically increasing function must fail
since the emergence of specific inhomogeneity scales due to perturbations at
a specific cosmic epoch is a discrete scenario. In view of hierarchically
nested discrete structure scales (galaxies, clusters of galaxies)
gravitational entropy as measure of the degree of inhomogeneity
in the universe is supposed to increase stepwise at each higher order
merging process. In addition, we require that a suitable gravitational
entropy representation should correspond in the black hole limit
to the general second law (GSL) \cite{Bekenstein73,Bekenstein94}
for the total entropy of the system. A master arrow of time is expected to
point in the direction of discrete increasing gravitational entropy
evolution as manifestation of irreversibility.

\section{Black hole entropy and bound systems}

The GSL of black hole dynamics provides the total entropy for systems
containing black holes S$^{tot}$ as balance between the black hole entropy $%
S^{int}$ proportional to the area of the event horizon $A$ and the entropy
of the surrounding matter $S^{ext}$ as

\begin{equation}
S^{tot}=S^{int}+S^{ext}=\frac{A}{l_{P}^{2}}+S^{ext}  \label{1}
\end{equation}

where $l_{P}$ is Planck's length and the requirement $\delta S^{tot}\geq 0$
for all processes is widely believed (we stress the functional dependence
and neglect factors of the order of unity, where appropriate; entropy is
measured in natural units). By dumping matter into a black hole any loss of
information or decrease of degrees of freedom in the outside world is is
stored in the event horizon. The existence of an upper bound for the entropy
or information capacity of any object of total energy $E$ and maximal size
$r$ was suggested by Bekenstein \cite{Bekenstein81,Bekenstein94}

\begin{equation}
S_{B}=\frac{Er}{\hbar c}=\frac{r_{g}r}{l_{P}^{2}}  \label{2}
\end{equation}

where $r_{g}$ is the gravitational radius.
Weaker bounds were proposed from the holographic principle \cite%
{Hooft93,Susskind95}, which suggests that the degrees
of freedom of a spatial region reside in the boundary, and were
illuminated in view of quantum information theory \cite%
{Bekenstein01} and cosmological implications \cite%
{Kaloper99,Easther99,Bak00}. Estimates of the maximal gravitational entropy
for the visible universe within a Hubble radius are found from the area of
the horizon $A_{H}$ as $S\sim (A_{H}/l_{P})^{2}\sim 10^{122}$
\cite{Kaloper99,Veneziano99,Barrow99} and we note that in de Sitter space also
the cosmological constant $\Lambda$ is quantized in terms of $l_{P}$ as
$S=N=(l_{P}^{2}\Lambda )^{-1}\sim 10^{122}$, constraint by the
Fiedmann-Robertson-Walker cosmology \cite{Garattini00}.

We rewrite the GSL for an ensemble of $N$ equal black holes of
mass $m$ localized in a universe of mass $M$ and size $R$ subject
to critical density with $M\sim R$ as

\begin{equation}
S^{tot}=NS^{int}+S^{ext}=N(\frac{r_{g}}{l_{p}})^{2}+(\frac{R}{r_{g}}%
)^{2}=Nn^{2}+N^{2}  \label{3}
\end{equation}

Equation (\ref{3}) provides the link to ordinary bound systems as outlined below
and the entropy bound (\ref{2}) is recovered for $r_{g}\rightarrow l_{p}$ or $%
r_{g}\rightarrow R$ with $(R/l_{p})^{2}\gg 1.$ Here consistency requires
that the external entropy contribution is measured in terms of black hole
units, the scale available in the system $R,$ rather than in Planck
units. If spacing of $R$ would be performed in terms of $l_{P}$ then the
maximum information capacity $(R/l_{P})^{2}$ of the system would be stored
in $S^{ext}$ only, a contradiction to the information capacity of the
individual black holes and the GSL. We argue therefore that the area spacing
of closed systems must be performed with respect to available scales of
closed subsystems. 

The maximum information a system can hold knowing in detail it's
configuration corresponds to the maximum entropy when knowing nothing about
it's internal states \cite{Bekenstein01}. Hence, gravitational entropy of a closed
system can be understood in view of a surface hiding the information
content of the internal region of spacetime where it's value measures the
missing information. We generalize this interpretation normally attributed
to black hole entropy to ordinary gravitationally bound systems. For
instance, the exact configuration of a globular cluster with respect to it's
gravitating quanta, the stars as next lower order subsystems, can be
identified from the outside world - an observer located in another cluster -
as lack of knowledge of the internal configuration. According to the
correspondence between information and entropy we adopt this as the
gravitational entropy contribution of the inhomogeneity cluster with regard
to the subsystems stars. On the same level we may refer to entropy in a
space region as to a quantity representing the degrees of freedom within
this region \cite{Brustein00a} where a gravitationally bound $N$-body system
is subject to $N(N-1)/2$ constraints represented by the links.
'Gravitational information' flows along all links between any pair of
constituents of the cluster and by knowing all links we would account also
for the missing information within the system. In particular, this view
reflects the causal set context where links are the building blocks of the
system and knowledge of the links between the elements of the causet is
equivalent to the knowledge of the whole causal set \cite%
{Rideout00,Sorkin00,Dou01}. Counting links with respect to the horizon was
found to be proportional to the horizon area in the black hole context \cite%
{Dou01}. But what are closed systems in a gravitational sense if we do not
refer to black holes? We define a gravitationally bound system only by the
number of links representing the interaction between any pair of subsystems,
equivalent to the number of constraints characterizing the degrees of
freedom of the system. Tracing closed systems hierarchically by links
between substructures was recognized by Leubner \cite{Leubner00a,Leubner00b}
to result in a unique configuration when applying naturally involved extremal
conditions constraining the gravitational entropy evolution due to formation
of structure in the universe \cite{Leubner00c}.

Consider now an ensemble of $N_{i}$ uniformly distributed structures of
species $(i)$ and mass $m_{i}$ in a universe, stars or galaxies for instance,
subject to gravitational interaction only. Let the system be rearranged due
to density perturbations into $N_{i+1}$ clusters of mass $m_{i+1}$, globular
star clusters or clusters of galaxies denoted as species $(i+1)$ of equal
richness $n_{i}$ in view of their next lower order substructures $(i)$. 
Since each member of a gravitationally bound system is affected by all other
constituents I propose the number of links $n_{i}(n_{i}-1)/2$ between all
members of a cluster to serve as measure of the gravitational entropy of the bound
$n_{i}-$ body system. The redistribution into $N_{i+1}$ clusters causes a
next higher order gravitational interaction between these systems as new
units represented by $N_{i+1}(N_{i+1}-1)/2$ links. Consequently, the entire
system is subject to a natural separation between internal and external
degrees of freedom. This is just what nature demonstrates since mutual
interaction of substructures of spatially separated bound macrosystems
is not realized. In this context the total gravitational entropy
contribution $\delta S_{i+1}$  as measure of the gain of inhomogeneity
within the two level system is provided by

\begin{equation}
\delta S_{i+1}=N_{i+1}n_{i}(n_{i}-1)/2+N_{i+1}(N_{i+1}-1)/2  \label{4}
\end{equation}

Applying large occupation numbers by $n_{i},N_{i+1}\gg 1$ and introducing a
mass $M$ of the entire system under consideration yields with $m_{i+1}=n_{i}m_{i}$

\begin{equation}
\delta S_{i+1}\sim N_{i+1}n_{i}^{2}+N_{i+1}^{2}\sim
N_{i+1}(m_{i+1}/m_{i})^{2}+(M/m_{i+1})^{2}  \label{5}
\end{equation}

a quantity proportional to the mass squared as required with regard to the
black hole limit where equation (\ref{3}) is recovered. 

\section{Hierarchy of structure scales}

Upon generalization a hierarchically growing clustering process \cite%
{Anderberg73}, constraint by elementary grouping
principles, can be formulated in a universe $\{G_{0}\}$, filled initially by
a specific number of particles $\{g_{0}\},$ see Figure 1. Let a clustering
procedure with regard to a certain structure level $(i)$ result in a
sequence of hierarchically nested sets of higher order clusters $(...\left\{
G_{i-1}\right\} ,\left\{ G_{i}\right\} ,\left\{ G_{i+1}\right\} ,\left\{
G_{i+2}\right\} ,...)$ where the members $G_{i}$\ of any specific level
admit equal richness, a constraint discussed below. Substructures $G_{i}$
sequentially $(n-$ times$)$ merge into higher order systems $G_{i+n}$
finally approaching the root, a universe identified as cluster with one
element. The two fundamental questions to be answered are: $(i)$ do
clusters of a common structure level admit nearly equal richness with regard
to their members, which raises the question of the emergence of equal structure
scales at the same cosmic time; and if confirmed: $(ii)$ how many
substructures merge in average forming sets of nearly equal occupation
numbers of a next higher order structure level, which raises the question of the
richness of equal structure scales with regard to their building blocks.

\begin{figure}[htb]
\centering
\includegraphics[height=2.5in]{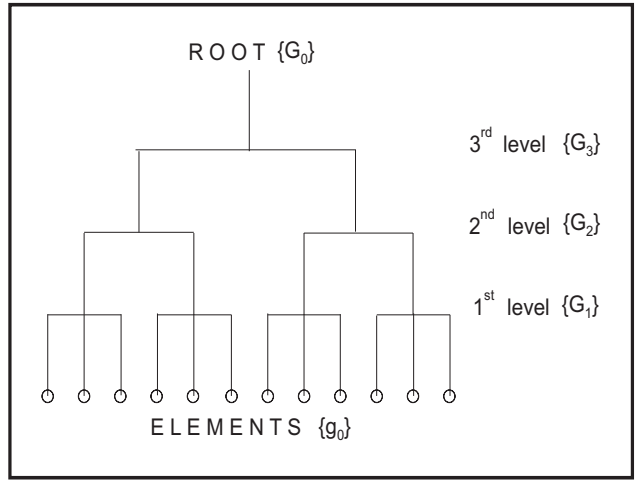}
    \caption{A hierarchical tree with 3 structure levels between the ground state
     and the root where the members of a common level admit equal richness.}
    \label{Figure1}
\end{figure}

In view of the unknown occupation numbers we verify the only
feasible configuration where the gravitational entropy increase within
any two levels of cluster formation $(i\rightarrow i+1)$ approaches
successively at each higher order merging process an extremum:

$(i)$ the sum of internal contributions $\delta
S_{i+1}^{int}=N_{i+1}S_{i}^{int}\sim N_{i+1}n_{i}^{2}$ is subject to
a {\it minimum} if all clusters $\left\{G_{i+1}\right\}$
have the same occupation number $n_{i}$ of elements
$\left\{ G_{i}\right\}$. This follows immediately from the quadratic
appearance of $n_{i}$ denoting the links between the elements together with
particle conservation with regard to $N_{i}$ and yields the equal richness
condition for bound systems belonging to the same structure level as

\begin{equation}
N_{i}=N_{i+1}n_{i}  \label{6}
\end{equation}
\qquad \qquad

wherefore the assumption in Figure 1 is justified.

$(ii)$ upon generalizing equation (\ref{4}) in view of a nested hierarchy of
structure levels the total entropy gain $\delta S_{i+1}$ at any transition
from some level $(i)$ to $(i+1)$ is defined by the condition

\begin{equation}
\delta S_{i+1}=N_{i+1}n_{i}(n_{i}-1)/2+N_{i+1}(N_{i+1}-1)/2\Longrightarrow
extremum  \label{7}
\end{equation}

again naturally satisfying a {\it minimum}. Inserting from equation
(\ref{6}) for $n_{i}$ where $N_{i}$ characterizes the prearranged structure
level $(i),$ and is therefore a frozen in constant of the system with regard
to higher order merging processes, we can solve for $N_{i+1}$ to arrive at a
simple recurrence condition for the quantity of identical new inhomogeneities
at level ($i+1)$ as

\begin{equation}
N_{i+1}=(N_{i}^{2}/2)^{1/3}  \label{8}
\end{equation}

where the limit of large occupation numbers $N_{i+1}^{3}>>N_{i+1}^{2}$ was used.
Relation (\ref{8}) can be written with regard to (\ref{6}) also in terms of
the individual cluster occupation number as $n_{i+1}=n_{i}^{2/3}$.
 
As natural ingredient of the entropy concept condition (\ref{7})
{\it minimizes} the total {\it increase} in gravitational entropy due to a
merging process from any level $i$ to level $i+1$. In other words, as
consequence of the separation this condition minimizes the sum of all
internal constraints with respect to the microsystems (building blocks) of
individual clusters and the external constraints between all clusters,
driving the entire environment into a state of highest degree of
autonomy. Any other configuration with
respect to both directions (many new systems with few members / few new
systems with many members where the limit of only one system where all
particles are linked reproduces to the original system) results in a
distribution of increased gravitational coupling between equal
members, equivalent to enhanced gravitational entropy contribution due to
clumping of structure from level $i$ to level $i+1$. The entropy gain is
successively reduced at each higher order gravitational merging process and
saturates. These are the the key issues of conditions (\ref{6}), (\ref{7}) and
(\ref{8}), respectively.

Let me introduce now physical observables and define an upper mass bound by
$M=N_{i}m_{i},$ equivalently applicable at any structure level $i+n$. Upon
substitution into equation (\ref{8}) with regard to the proper indices, the
recurrence relation for the mass scales of inhomogeneities permitted by the
entropy constraint reads

\begin{equation}
m_{i+1}=(2m_{i}^{2}M)^{1/3}  \label{9}
\end{equation}

Approximate radii $r_{i}$ of structure scales may be found from a basic
principle of statistics for the mean error of spatial uncertainties of an
ensemble of $n_{i}$ particles within a volume of dimension $r_{i+1}$ from
$r_{i}=r_{i+1}/\sqrt{n_{i}}$, a relation interpretable also as area
quantization in view of the holographic principle. This yields an invariant
effective for any fundamental structure scale as

\begin{equation}
\frac{m_{i}}{r_{i}^{2}}=\frac{m_{i+1}}{r_{i+1}^{2}}=...=\frac{M}{R^{2}}%
=\Sigma _{0}=const.  \label{10}
\end{equation}

and we note already here that significant support for a functional
dependence $m\propto r^{2},$ a constant surface density $\Sigma _{0}$ for
astrophysical objects was found on observational grounds, see section 4.
After combining relations (\ref{9}) and (\ref{10}) the appropriate
condition for the spatial dimensions of inhomogeneities subject to the
entropy constraint reads

\begin{equation}
r_{i+1}=(\sqrt{2}r_{i}^{2}R)^{1/3}  \label{11}
\end{equation}

Finally, in view of astrophysical cluster analysis \cite{Xu00}\ we
characterize by means of the richness $n_{i}$ a cluster halo by
the mean separation of neighbor clusters in recurrence notation as

\begin{equation}
d_{i}=\frac{2r_{i+1}}{n_{i}^{1/3}}  \label{12}
\end{equation}

\section{Observational test and discussion}

A simple set of recurrence relations determines a global sequence of
inhomogeneity scales in a predefined Hubble volume from gravitational
entropy restrictions. Type Ia supernovae and microwave background observations
support presently a flat, accelerating universe of critical density where the
density parameter $\Omega $ splits as $\Omega =0.7\Omega _{q}+0.3\Omega
_{m}=1$ into a dark energy (quintessence) component and a matter
contribution where a Hubble parameter of $H_{0}=70\ km\ \sec ^{-1}Mpc^{-1}$
is typically favored \cite{Bludman01}. For an Einstein de Sitter universe
the set of equations determining the sequential growth of structure scales can be
solved by introducing Hubble's parameter only. This is a consequence of a
natural ingredient of the proposed entropy concept since Planck's length,
defining as lowest spatial bound the limit of information capacity at
$r_{0}=l_{P},$ provides a starting value, which generates a sequence that
accurately reproduces observations on astrophysical scales. The constant
surface density on the right hand side of equation (\ref{10}) is of the order
of unity and available from the critical density wherefore also the starting value
of the mass sequence is found as $m_{0}=l_{P}^{2}\Sigma _{0}$. With regard to
equation (\ref{10}) the quantity $S_{0}=N_{0}=M/m_{0}=(R/l_{P})^{2}=
c^5/(\hbar H_{0}^{2}G)=6.7\times 10^{121}$ is obtained as initial condition for
equation (\ref{8}) and corresponds to the current value of the
Bekenstein-Hawking entropy of the universe inside a Hubble radius, identical
to the quantization condition for the cosmological constant $\Lambda$, 
introduced in section 2. Furthermore, with the use of Plank's mass $m_{P}$ a
recently proposed bound on $H_{0}$ from the largest geometric entropy per
Hubble volume is reproduced as $H_{0}\leq m_{P}c^2/\hbar \sqrt{N_{0}}$ \cite
{Brustein00b}. According to equation (\ref{7}) the proportionality $S\sim N^{2}$ appears
to generate a contradiction to the maximum entropy content within a Hubble volume with
regard to the first two structure levels. This is resolved naturally by applying the
term 'gravitational entropy' only for gravitational interaction with the restriction to
mass scales above Planck's mass $m>m_{P}$, or $N<\sqrt{N_{0}}$ and we keep below
the required proportionality of radiation entropy $S\sim N,$ which reflects
also the transition from a radiation dominated to a matter dominated universe.

\medskip \bigskip

\begin{center}
Table 1: Scaling properties of fundamental structures\\[0pt]
\bigskip 
\begin{tabular}{|c|cccc|c|}
\hline
\textbf{class} & $\mathbf{m}_{i}\mathbf{[g]}$ & $\mathbf{r}_{i}\mathbf{[cm]}$
& $\mathbf{d}_{i}\mathbf{[cm]}$ & $\mathbf{N}_{i}$ & \textbf{object} \\ 
\hline
$g_{0}$ & 2.7$\times $10$^{-66}$ & 1.6$\times 10^{-33}$ & 2.0$\times 10^{-26}
$ & 6.7$\times 10^{121}$ & Planck scale \\ 
$G_{1}$ & 1.3$\times 10^{-25}$ & 3.6$\times 10^{-13}$ & 2.4$\times 10^{-8}$
& 1.3$\times 10^{81}$ & hadronic matter \\ 
$G_{2}$ & 1.9$\times 10^{2}$ & 1.4$\times 10^{1}$ & 2.8$\times 10^{4}$ & 9.5$%
\times 10^{53}$ & condensed matter \\ 
$G_{3}$ & 2.3$\times 10^{20}$ & 1.5$\times 10^{10}$ & 3.1$\times 10^{12}$ & 
7.7$\times 10^{35}$ & planetesimals \\ 
$G_{4}$ & 2.7$\times 10^{32}$ & 1.6$\times 10^{16}$ & 7.1$\times 10^{17}$ & 
6.6$\times 10^{23}$ & stellar systems \\ 
$G_{5}$ & 2.9$\times 10^{40}$ & 1.7$\times 10^{20}$ & 2.7$\times 10^{21}$ & 
6.0$\times 10^{15}$ & globular clusters \\ 
$G_{6}$ & 6.7$\times 10^{45}$ & 8.1$\times 10^{22}$ & 6.4$\times 10^{23}$ & 
2.6$\times 10^{10}$ & galaxies \\ 
$G_{7}$ & 2.5$\times 10^{49}$ & 5.0$\times 10^{24}$ & 2.5$\times 10^{25}$ & 
7.0$\times 10^{6}$ & galaxy clusters \\ 
$G_{8}$ & 6.1$\times 10^{51}$ & 7.7$\times 10^{25}$ & 2.8$\times 10^{26}$ & 
2.9$\times 10^{4}$ & superclusters \\ 
-- & -- & -- & -- & -- & -- \\ 
$G_{0}$ & 1.8$\times 10^{56}$ & 1.3$\times 10^{28}$ & 1.3$\times 10^{28}$ & 1
& the universe \\ \hline
\end{tabular}
\end{center}

\medskip \bigskip

Based on $H_0$ some instructive parameters of the predicted global sequence of 
structure scales are presented in Table 1 and illuminated in Fig. 2. When testing
a global structure quantization subject to ten discrete inhomogeneity scales within
120 orders of magnitude the question of mixing up objects with and without dark
matter seems to be of no relevance. On the other hand, comparing the derived
astrophysical mass scales with observations infers that the provided values must
be identified as total mass of the specific inhomogeneity scale, including
dark matter contributions on astrophysical scales. In this respect it is also
possible to deduce from the results of the gravitational entropy concept and the
knowledge of luminous matter the amount of dark matter mass contributions
of a certain astrophysical structure scale where table 1 also suggests
to identify $r_{i}$ with the luminous matter distribution and $d_{i}$
with the halo dimensions. Moreover, the solution should be regarded as
representative at timescales directly after relaxation and virialization
of a specific structure disregarding subsequent evolutionary aspects as
discussed recently in view of star clusters \cite{Burkert00}. Only a brief
outline with basic referencing for readers not familiar with the status regarding
all different structure scales can be provided here to illuminate the proposed
entropy concept in view of theoretical and observational evidence.

\begin{figure}[htb]
\centering
    \includegraphics[height=4in]{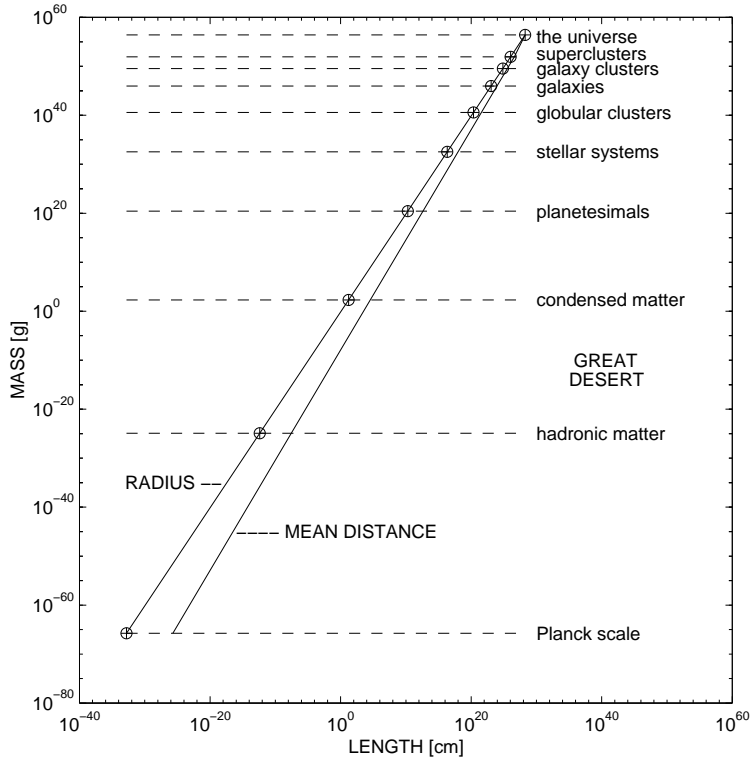}
    \caption{A schematic illustration of the hierarchy of structure scales. Radius
and mean distance are obtained from equations (11) and (12).}
    \label{Figure2}
\end{figure}

It is reasonable to assign the {\bf ground state} ${\bf g}_{0}$ to a
gravitational background \cite{Ben-Jacob99} to which any bound system is
coupled. Constraints on the mass are available from gravitational wave
observations \cite{Will98} and arguments were provided that\ gravity can
appear due to polarization of instantons in the SO(4) gauge theory where
their radius is comparable to Planck's length $l_{P}=r_0$ in table 1 \cite{Kuchiev98}.
Recent attempts to explain the non-vanishing small value of the vacuum
energy density favor a quintessence background as alternative to a
cosmological constant \cite{Wang00,Leubner00c,Bludman01}. This is supported by
the starting value of the mass sequence $m_{0}$, which coincides
with the acquired field mass necessary to dominate the current
energy density of the universe \cite{Frieman95}. The transition from the ground
state to hadronic matter reflects the ratio of strong to gravitational
interaction where ground state hadrons as representatives of the
{\bf elementary particle physics scale} $G_{1}$ are subject to a quark
confinement length $r_{1}$. Introducing by $\lambda_{n}=r_{1}$
the neutron Compton wavelength, Dirac's hypothesis for the particle
number in the universe $N_{1}=(R/\lambda _{n})^{2}$ \cite{Dirac37}
is a consequence of the proposed approach and also the popular
Weinberg coincidence \cite{Weinberg72} for the pion mass $m_{\pi }^{3}\simeq
\hbar^2 H_{0}/cG$ turns out as implicit content. Hence, the mystery of large
numbers appears as simple content of the proposed global quantization
concept. Proceeding through a 'great desert', predicted by force
unification theories, to the structure scale $G_{2}$ a mass density of the order
of unity is predicted. In view of an assignment to a specific bound system
evidently a high degree of structure variety is today available for
'clusters of atoms' denoted here as {\bf condensed matter}. Interestingly,
upon calculating the mean separation of the constituents of $G_{1},$ protons
for instance, with respect to the characteristic cluster domain $r_{2},$ from
(\ref{12}) a value $d_{1}\simeq 10^{-8}cm$ of the order of Bohr's radius
is found. This implies in conjunction with the entropy constraint
the existence of a representative structure scale with a mass
density of the order of unity, bound on atomic dimensions where prestellar
grain may serve as possible candidate in view of the evolutionary history
of the universe \cite{Milani94,Benz00}.

On intermediate scales{\bf \ planetesimals }$G_{3}$, comets and asteroids
play a key role in theories of the evolution of the solar system
\cite{Milani94,Benz00}. Within the formation of a massive extended protoplanetary
disk, the development of a large number of comets and planetesimals with
individual masses of the order of $m_{3}\simeq 10^{20}g$ provides the link
between the phase of condensed matter $G_{2}$ and stellar systems $G_{4}$. A
protoplanetary disk of solar mass $m_{4}\simeq 10^{33}g$ and a radius of
$10^{14}-10^{15}cm$ can contain a number of $n_{3}\simeq 10^{13}$
planetesimals. Referring to evolutionary theories, a
marginally unstable cloud of solar mass $m_{\odot }\simeq 10^{33}g,$
destined to form a {\bf stellar system} $G_{4}$\ has a radius of the order
of $10^{17}cm,$ where the disk has a limiting lower mass of typically
$m\simeq 0.05m_{\odot }$. Table 1 reflects all values
sufficiently and is supported observationally also by the orbital
characteristics of solar system comets and Kuiper populations defining the
edge of the bound solar system \cite{Milani94,Allen00}. Next, from the
entropy constraint a structure scale of the dimension of {\bf globular clusters}
$G_{5}$ is predicted where observed radii can be averaged at about
$35pc\simeq 1\times 10^{20}cm$ reaching up to $100pc$ surrounded by large dark
matter haloes \cite{Heggie96}. Remarkably, old globular clusters were well fitted by a
functional form $m\sim r^{2}$ for their upper mass where a discrepancy with respect
to young clusters was suggested to indicate different formation histories or
later evolutionary effects \cite{Burkert00}. This recent observational
development where a cutoff mass $m_{c}$ was found in the range
$10^{6}M_{\odot }\leq m_{c}\leq 5\times 10^{6}M_{\odot }$ clearly
supports the constant surface density (\ref{10}) as limiting condition,
indicating also that the presented results are representative for timescales
just after the formation of a specific structure, disregarding evolutionary
changes. The outermost halo-globulars of the galaxy are found at about
$100kpc$ from the galactic center indicating the galactic boundary, a value
provided also from dark matter halo studies for galaxies \cite{Lang99}.

At large astrophysical scales, the optically visible cores of representative
members of {\bf galaxies} $G_{6}$ have mean radii $r_{6}$ in the range of 20 kpc
with a dominant dark matter mass of 2$\times 10^{12}m_{\odot }$.
In average the halo mass distribution extends to about 200 kpc \cite%
{Lang99,Klypin01}, wherefore galaxies can be identified safely as
structure level 6 of the hierarchy. {\bf Clusters of galaxies} $G_{7}$ are
the largest virialized inhomogeneities in the universe \cite{Xu00} with a
total mass estimated up to$\ 10^{16}m_{\odot }\simeq 10^{49}g$ and the
dominant dark matter is distributed within a halo of about $5Mpc\simeq
10^{25}cm.$ Galaxy cluster formation is dated after the evolution of
galaxies supporting the hierarchical scenario \cite{Fabian92} where already
early studies have indicated a constant surface density from N-body
simulations \cite{Kashlinsky83,West89}. Observations of the
distribution of {\bf superclusters }$G_{8}$ identify a network structure of
scales up to $(100-150)$ $Mpc\simeq $ ($3-4.5)\times 10^{26}cm,$ irregular
in shape since not fully virialized. The supercluster network is surrounded
by low density regions of similar scale, suggesting a cellular structure of
the universe at such dimensions. Superclusters contain a fraction of $20-100$
clusters with a mean mass estimated in the range of $m_{8}\simeq 5\times
10^{50}g$ \cite{Fabian92} and a separation scale of the order of their
proper size implying that the predictions of the gravitational entropy
approach are sufficiently supported by observations also on largest scales
yet observed.

In further steps, the sequence converges to the {\bf universe} $G_{0}$
as $N_{i}\Rightarrow 1,$ where equation (\ref{7}) must be solved exactly and
where the definition of a virialized bound system as unique cluster is not
applicable anymore. The gravitational entropy constraint predicts the
emergence of ten fundamental inhomogeneity scales in the universe
including the groundstate ($m_{0},r_{0})$ and a universe ($M,R)$ as bounds,
a result determined between $10^{122}\geq N_{i}\geq 1$.

\begin{figure}[htb]
\centering
    \includegraphics[height=2.5in]{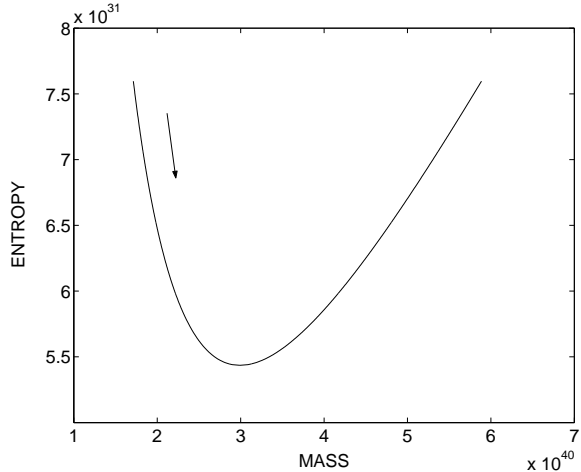}
    \caption{The mass evolution for globular clusters in view of the minimum
             gravitational entropy condition.}
    \label{Figure3}
\end{figure}

The proposed gravitational entropy concept extracts
the only configuration of a system where the increase in entropy or
information capacity due to higher order clumping of inhomogeneities is
minimized. Fig. 3 demonstrates the functional entropy-mass dependence,
as obtained from (\ref{7}) after substituting for the mass, with regard
to the evolution of the globular cluster scale as example. Due to
clumping of structure the entropy of the preceeding level drops down as
indicated by the arrow reaching the minimum in a relaxed state at a cluster
cutoff mass $m_{5}=$ $2.9\times 10^{40}g$, supported by observations, where
further mass aggregation would result again in a state of increased entropy.
A minimum growth of entropy guarantees a configuration of maximum
possible inhomogeneity at each structure level. Adding all entropy
contributions of all degrees up to level $i+1$ yields

\begin{equation}
S^{tot}=\mathrel{\mathop{\sum }\limits_{i}}
\left[ N_{i+1}n_{i}(n_{i}-1)+N_{i+1}(N_{i+1}-1)\right] ,\hspace{1cm}\delta
S^{tot}>0
\end{equation}

indicating that gravitational entropy is increasing in a quantized manner at
each higher order inhomogeneity level and defines thermodynamically consistent
a gravitational master arrow of time. This one-way character allows spacetime
to develop into states of increasingly nested higher order structure scales.

Increasing inhomogeneity due to gravitational clumping reflects growth of
gravitational entropy in an evolving universe. In this respect a measure of
gravitational entropy was introduced reproducing the GSL of black hole
dynamics in the limit. The gain of gravitational entropy at each higher
order merging process is suggested to result from an extremal condition,
requiring a minimum increase of entropy. The proposed approach reproduces the
observed global inhomogeneity scales where astrophysical scales, subject to
gravitational interaction, are linked to the particle physics and Planck's scale
within one unique concept. Thermodynamic equilibrium at the big bang requires the
entropy to approach about it's maximum value $S_{\max }$. Since any
gravitational contribution at some level $m_{i}>m_{P}$ is subject to 
$S_{i}\ll S_{\max }$ the entropy paradox appears to be resolved. Consistent
with the thermodynamic view a gravitational master arrow of time can be
defined that points in the direction of increasing entropy or
inhomogeneity, generating a one-way character of the future. The underlying
gravitational entropy concept implies in view of fundamental structure scales
that we live in a universe of maximum autonomy.

%%%%%%%%%%%%%%%%%%%%%%%%%%%%%%%%%%%%%%%%%%%%%%%%%%%%%%%%%%%%%%%%%%%%%%
%%
%%   use this format to include an .eps figure into your paper
%%
%\begin{figure}[htb]
%    \centering
%    \includegraphics[height=1.5in]{cosmo.eps}
%    \caption{Figure caption.}
%    \label{fig:cosmo}
%\end{figure}
%%%%%%%%%%%%%%%%%%%%%%%%%%%%%%%%%%%%%%%%%%%%%%%%%%%%%%%%%%%%%%%%%%%%%%%%

%\Acknowledgements

\end{document}